\documentclass{iopart}
\usepackage{iopams,amsfonts,amssymb,amsthm} 
\usepackage{geometry}                
\usepackage{graphicx}
\usepackage{rawfonts}
\usepackage{amssymb}
\usepackage{epstopdf}
\usepackage[usenames,dvipsnames,svgnames]{xcolor}
\input prepictex.tex
\input pictex.tex
\input postpictex.tex

\begin{document}

\newtheorem{theo}{Theorem}
\newtheorem{lemm}{Lemma}
\newtheorem{corr}{Corollary}
\newtheorem{defn}{Definition}
\newtheorem{remark}{Remark}

\def\Pr{\noindent \emph{Proof: }}
\def\qed{$\Box$}

\title[Self-avoiding walks adsorbed at a surface and subject to a 
force] {Self-avoiding walks adsorbed at a surface and subject to a 
force\footnote[3]{\textsf{Dedicated to Tony Guttmann on the occasion of his 70th birthday}}}
\author{E J Janse van Rensburg$^1$\footnote[1]{\textsf{rensburg@yorku.ca}}
 and S G Whittington$^2$\footnote[2]{\textsf{swhittin@chem.utoronto.ca}}}
\address{$^1$ Department of Mathematics, York University, Toronto, Canada}
\address{$^2$ Department of Chemistry, University of Toronto, Toronto, Canada}
\begin{abstract}
We consider self-avoiding walks terminally attached to an impenetrable surface at which
they can adsorb.  We call the vertices farthest away from this plane the \emph{top vertices}
and we consider applying a force at the plane containing the top vertices.  This force 
can be directed away from the adsorbing surface or towards it.  In both cases we prove that 
the free energy (in the thermodynamic limit) is identical to the free energy when the force
is applied at the last vertex.  This means that the criterion determining the critical 
force - temperature curve is identical for the two ways in which the force is applied and the response
to pushing the walk is also the same in the two cases.

\end{abstract}

\submitto{J Phys A}
\maketitle

\section{Introduction}
\label{sec:Introduction}

The adsorption of polymers at surfaces is a well established subject that still attracts interest \cite{Rensburg2015}.  
With the advent of micro manipulation techniques such as atomic force microscopy (AFM) and optical tweezers it has
become possible to pull adsorbed polymers off a surface \cite{Haupt1999,Zhang2003}.  That is, the polymer
can be desorbed by a mechanical force.  In principle one can obtain the force-extension 
curve or the temperature dependence of the critical force for desorption.

\begin{figure}[h]
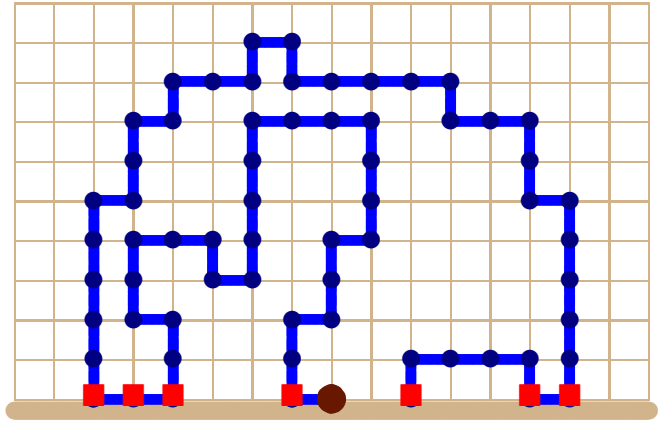

\beginpicture
\setcoordinatesystem units <1.5pt,1.5pt>
\setplotarea x from -140 to 110, y from -10 to 100
\setplotarea x from -80 to 80, y from 0 to 100

\color{Tan}
\grid 16 10 
\color{Tan}
\setplotsymbol ({\LARGE$\bullet$})
\plot -80 -3 80 -3 /

\setplotsymbol ({$\bullet$})
\color{Blue}
\plot -60 0 -50 0 -40 0 -40 10 -40 20 -50 20 -50 30 -50 40 -40 40 -30 40 -30 30 
-20 30 -20 40 -20 50 -20 60 -20 70 -10 70 0 70 10 70 10 60 10 50 10 40 0 40 0 30 
0 20 -10 20 -10 10 -10 0 -10 0 0 0 /
\plot 20 0 20 10 30 10 40 10  50 10 
50 0 60 0 60 10 60 20 60 30 60 40 60 50 50 50 50 60 50 70 40 70 30 70 30 80 20 80
10 80 0 80 -10 80 -10 90 -20 90 -20 80 -30 80 -40 80 -40 70 -50 70 -50 60 -50 50 -60 50
-60 40 -60 30 -60 20 -60 10 -60 0   /
\color{NavyBlue}
\multiput {\LARGE$\bullet$} at 
-60 0 -50 0 -40 0 -40 10 -40 20 -50 20 -50 30 -50 40 -40 40 -30 40 -30 30 
-20 30 -20 40 -20 50 -20 60 -20 70 -10 70 0 70 10 70 10 60 10 50 10 40 0 40 0 30 
0 20 -10 20 -10 10 -10 0 -10 0 0 0   20 0 20 10 30 10 40 10  50 10 
50 0 60 0 60 10 60 20 60 30 60 40 60 50 50 50 50 60 50 70 40 70 30 70 30 80 20 80
10 80 0 80 -10 80 -10 90 -20 90 -20 80 -30 80 -40 80 -40 70 -50 70 -50 60 -50 50 -60 50
-60 40 -60 30 -60 20 -60 10 -60 0   /

\color{red}
\multiput {\large$\blacksquare$} at -60 1 -50 1 -40 1 -10 1 20 1 50 1 60 1 /

\color{Sepia}
\put {$\huge\bullet$} at 0 0
\circulararc 360 degrees from 2.3 0 center at 0 0
\color{black} \normalcolor
\endpicture

\caption{A positive adsorbing walk.  Visits to the distinguished hyperplane are weighted by $a$ and denoted by
$\color{red}\blacksquare$.}
\label{fig:poswalk}
\end{figure}

A natural model is a self-avoiding walk in a half-space, interacting with the line or plane defining 
the half-space.  If we consider the $d$-dimensional hypercubic lattice, with coordinate system 
$(x_1,x_2,\dots x_d)$, we can consider a self-avoiding walk, starting at the origin and having all
vertices with non-negative $x_d$-coordinate.  These are \emph{positive walks}.  See Figure \ref{fig:poswalk}.
Each vertex that is in the hyperplane $x_d=0$ is a 
\emph{visit} and the walk can be weighted according to the number of visits to model the 
adsorption process.  Without a force, the adsorption process is quite well understood and it is
known that, in the infinite
size limit, there is a phase transition \cite{HTW} from the adsorbed to the desorbed state as 
the temperature is increased.  If a force is applied to desorb the polymer, the problem has been
investigated numerically by several groups \cite{Guttmann2014,Krawczyk2005,Mishra2005}
and some rigorous results are also available \cite{Guttmann2014,Rensburg2013}.  In all of these 
papers  the force was applied, normal to the surface, at the last vertex of the walk.

Suppose that $c_n^+(v,h)$ is the number of $n$-step positive walks with $v+1$ visits and with the $x_d$-coordinate of the
$n$'th vertex equal to $h$.  Define the partition function 
\begin{equation}
C_n^+(a,y) = \sum_{v,h} c_n^+(v,h) a^v y^h
\end{equation}
where $a= e^{-\epsilon /k_BT}$, $y=e^{f/k_BT}$, $\epsilon$ is the energy associated with a 
visit, $f$ is the applied force, $k_B$ is Boltzmann's constant and $T$ is the absolute temperature.
It is known that the free energy $\psi^+(a,y)$ defined by the limit
\begin{equation}
\psi^+(a,y) = \lim_{n\to\infty}  {\scriptsize\frac{1}{n}} \log C_n^+(a,y)
\end{equation}
exists for all $a > 0$ \cite{Rensburg2013}.  If we set $y=1$ (turning off the force) we have the pure adsorption
problem and we write $\psi^+(a,1) = \kappa(a)$.    There is a critical value of $a$, $a_c > 1$, such that 
$\kappa(a) = \kappa(1) = \log \mu_d$ for $a \le a_c$ and $\kappa(a) > \log \mu_d$ for $a > a_c$ \cite{HTW}.
Here $\mu_d$ is the growth constant for self-avoiding walks on ${\mathbb Z}^d$ \cite{Hammersley1957,MadrasSlade}.
If we set $a=1$ so that there is no interaction with the surface ($\epsilon = 0$) we can write 
$\psi^+(1,y)=\lambda(y)$.   There is a transition to a ballistic phase 
at $y=y_c=1$ \cite{Beaton2015}.  See also \cite{IoffeVelenik} for related work.

If we look at the general case where both $a$ and $y$ are not necessarily equal to 1 then,
for $a \le a_c$ and $y \le 1$, $\psi^+(a,y) = \log \mu_d$ \cite{Rensburg2013}.  When 
$a > a_c$ and $y > 1$
\begin{equation}
\psi^+(a,y) = \max [\kappa(a), \lambda(y)]
\end{equation}
and there is a phase boundary, $y=y_c(a)$, in the $(a,y)$-plane separating
the adsorbed phase from the ballistic phase, determined by the solution of the 
equation $\kappa(a) = \lambda(y)$ \cite{Rensburg2013}.  The phase transition on crossing this 
boundary is first order \cite{Guttmann2014}.  

Without much loss of generality we can set $\epsilon = -1$ and switch to the force-temperature
plane.  When $d \ge 3$ the phase boundary $f=f_c(T)$ is re-entrant \cite{Rensburg2013,Krawczyk2005,Mishra2005}
but the critical force is a monotone decreasing function of the temperature 
when $d=2$ \cite{Guttmann2014,Mishra2005}.

All of the results discussed above are for the case where the force is applied at the last vertex of the walk.
In an AFM experiment, unless special precautions are taken, the tip might not contact the last monomer so it is 
interesting to enquire how robust the above results are.  Another case that has been considered is
to apply the force to vertices whose $x_d$-coordinate is largest.  One can think of having a 
confining plane at $x_d=s$ with at least one vertex in this plane, and then varying the value of
$s$.  The \textit{height} $s$ is the span of the walk in the $x_d$-direction, and $y$ is conjugate to this height.  

For the 
special case of $a=1$ this problem has been investigated \cite{GuttmannLawler,Rensburg2009,Rensburg2016}
and both numerical and rigorous results are available.  In a Monte Carlo
investigation in $d=3$ \cite{Rensburg2009} this case was compared to the situation where the force is 
applied at the last vertex and, for $y > 1$, the results are very similar even for modest values 
of $n$.  When $y < 1$ there are substantial differences between the two modes whereby the force 
is applied when $n$ is as large as 4000.  Recall that when $y < 1$, $f < 0$ and the walk is being
pushed towards the surface.  

For the two dimensional case Beaton \emph{et al} \cite{GuttmannLawler} have used SLE 
ideas to investigate the situation when $y < 1$.  They find unexpected stretched 
exponential terms leading to slow convergence to the limiting value
of the free energy.  The limiting free energy has been shown to be 
equal to $\lambda(y)$ for all $y > 0$, \emph{i.e.} both when the walk is being 
pulled and when it is being pushed \cite{Rensburg2016}.

In this paper we extend some of the results in \cite{Rensburg2016} to the case where $a \ne 1$ so
that the walk can be adsorbed at a surface and then pulled off by the second method of
applying the force, described above.  Our principle result is that the limiting free 
energy is the same for both ways of applying the force.

\section{Desorbing a self-avoiding walk by applying a force}
\label{sec:SAWpull}

We consider positive walks on ${\mathbb Z}^d$ where we keep track of the number of visits to the 
hyperplane $x_d=0$ and the span in the $x_d$-direction.  Let $c_n(v,s)$ be the number
of $n$-step positive walks in ${\mathbb Z}^d$  with $v+1$ visits to $x_d=0$ and with span in the $x_d$-direction
equal to $s$.  Define the partition function 
\begin{equation}
C_n(a,y) = \sum_{v,s} c_n(v,s) a^v y^s
\label{eqn:pfspan}
\end{equation}
where $a$ and $y$ are interpreted as described in the Introduction.  Clearly
\begin{equation}
\lim_{n\to\infty}  {\scriptsize\frac{1}{n}} \log C_n(a,1) = \psi^+(a,1) = \kappa(a)
\label{eqn:noforce}
\end{equation}
and it was proved in \cite{Rensburg2016} that
\begin{equation}
\lim_{n\to \infty}  {\scriptsize\frac{1}{n}} \log C_n(1,y) = \psi^+(1,y) = \lambda(y).
\label{eqn:nointeraction}
\end{equation}

We first consider the case when the walk is attracted to the surface and the force is pulling the 
walk off the surface, \emph{i.e.}   $\epsilon < 0$ and $f > 0$.

\begin{lemm}
When $a \ge 1$ and $y \ge 1$  
$$\liminf_{n\to\infty}  {\scriptsize\frac{1}{n}} \log C_n(a,y) \ge \max [\kappa (a), \lambda(y)].$$
\end{lemm}
\Pr From (\ref{eqn:pfspan}) it is clear that $C_n(a,y)$ is a monotone increasing function of 
both $a$ and $y$.  The Lemma then follows from (\ref{eqn:noforce}) and (\ref{eqn:nointeraction}).
\qed

We next turn to the issue of proving a suitable upper bound.   We define a \emph{loop} to be a positive walk
whose last vertex is in $x_d=0$ so that the walk starts and ends in this hyperplane.  Suppose that $L_n(a,y)$
is the partition function of $n$-step loops where $a$ is conjugate to the number of visits and $y$ is conjugate
to the span in the $x_d$-direction.  The partition function for loops with $(n+2)$ steps that leave $x_d=0$ 
at their first step and return for the first time at 
their $(n+2)$'th step is $ay L_n(1,y)$.  We say that a loop is \emph{unfolded in the $x_1$-direction} if the 
$x_1$-coordinate of every vertex is at least as large at that of the first vertex and the $x_1$-coordinate
of every vertex except the last vertex is strictly less than that of the last vertex.  We write
 $L_n^{\ddagger} (a,y)$
for the partition function of $n$-step unfolded loops.  Then $L_n^{\ddagger}(a,y) \le L_n(a,y) \le e^{O(\sqrt n)} L_n^{\ddagger}(a,y)$
\cite{HammersleyWelsh}.  It is known  \cite{HTW} that 
\begin{equation}
\lim_{n\to\infty}  {\scriptsize\frac{1}{n}} \log L_n(a,1) = \lim_{n\to\infty}  {\scriptsize\frac{1}{n}} \log L_n^{\ddagger}(a,1) = \kappa (a).
\end{equation}

\begin{figure}[t]
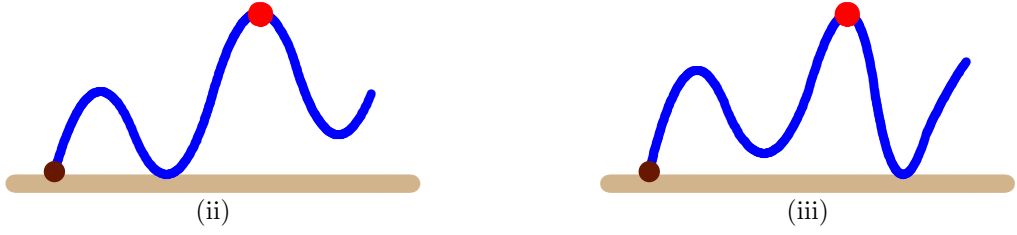

\beginpicture
\setcoordinatesystem units <1.5pt,1.5pt>
\setplotarea x from -75 to 50, y from -10 to 50
\setplotarea x from -50 to 50, y from -10 to 50

\color{black}
\put {(ii)} at 0 -10 

\color{Tan}
\setplotsymbol ({\LARGE$\bullet$})
\plot -50 -3 50 -3 /

\setquadratic
\color{blue}
\setplotsymbol ({\footnotesize$\bullet$})
\plot -40 0 -30 20 -20 10 -10 0 0 20 10 40 20 30 30 10 40 20  /
\color{red}
\put {\LARGE$\bullet$} at 12 40 
\circulararc 360 degrees from 14 40 center at 12 40 

\setlinear
\color{Sepia}
\put {\huge$\bullet$} at -40 0 
\setcoordinatesystem units <1.5pt,1.5pt> point at -150 0 
\setplotarea x from -50 to 50, y from -10 to 50

\color{black}
\put {(iii)} at 0 -10 

\color{Tan}
\setplotsymbol ({\LARGE$\bullet$})
\plot -50 -3 50 -3 /

\setquadratic
\color{blue}
\setplotsymbol ({\footnotesize$\bullet$})
\plot -40 0 -30 25 -20 15 -10 5 0 20 10 40 17 20 23 0 30 10 35 20 40 28  /
\color{red}
\put {\LARGE$\bullet$} at 10 40 
\circulararc 360 degrees from 12 40 center at 10 40 

\setlinear

\color{Sepia}
\put {\huge$\bullet$} at -40 0 

\color{black} \normalcolor
\endpicture

\caption{Cases (ii) and (iii) in the proof of lemma 2.  In case (ii) the walk does not return
to the adsorbing surface after it passes through its last vertex at maximum height,
and in case (iii) the walk passes through its last vertex at maximum height and then
returns at least once to the adsorbing surface.}
\label{fig:cases}
\end{figure}

\begin{lemm}
When $a \ge 1$ and $y \ge 1$  
$$\limsup_{n\to\infty}  {\scriptsize\frac{1}{n}} \log C_n(a,y) \le \max [\kappa (a), \lambda(y)].$$
\end{lemm}
\Pr Consider a positive walk with $n$ steps and 
span $s$.  Let $m$ be the index of the last vertex in the hyperplane $x_d=s$.  Then we have three cases to consider:
\begin{enumerate}
\item
$m=n$ and the walk is entirely in $x_d=0$.
\item
The walk does not return to $x_d=0$ after the $m$'th vertex.
\item
The walk returns to $x_d=0$ after the $m$'th vertex.
\end{enumerate}
See Figure \ref{fig:cases} for a sketch of cases (ii) and (iii).
For case (i) the free energy is $\kappa_{d-1} + \log a $ and this is bounded above by $\kappa (a)$.
For case (ii) suppose that the last vertex in $x_d=0$ is the $k$'th vertex.  Then
the partition function of these walks is bounded above by
$\sum_{0 \le k < n} C_k(a,1) C_{n-k}(1,y)$
and the corresponding generating function is 
\begin{equation}
G_1(a,y,t) = \sum_{k,n} C_k(a,1) C_{n-k}(1,y) t^n = \left(\sum_{n\ge 0} C_n(a,1)t^n \right)\left(\sum_{n\ge 1} C_n(1,y) t^n\right).
\end{equation}
$G_1$ is singular at $t=e^{-\kappa (a)}$ and at $e^{-\lambda (y)}$.
For case (iii) suppose that $k_1$ is the last visit  before the $m$'th vertex and that the walk returns to 
$x_d=0$ at the $k_2$'th vertex.  The partition function of these walks is bounded above by
$$\sum_{3\le k_2 \le n}\left(\sum_{0\le k_1\le k_2-3} C_{k_1}(a,1) a y   L_{k_2-k_1-2}(1,y) C_{n-k_2}(a,1) \right).$$
Note that the first and last edges in the loop are constrained to be 
normal to the adsorbing surface and the loop has no intermediate visits.  The partition function
of these loops is $ayL_{k_2-k_1-2}(1,y)$.
The corresponding generating function of the upper bound is given by
\begin{eqnarray}
G_2(a,y,t) & = & \sum_{k_1,k_2,n} C_{k_1}(a,1) a y L_{k_2-k_1-2}(1,y) C_{n-k_2}(a,1) t^n 
\nonumber \\
& = & a y t^{2}   \left(\sum_n  C_n(a,1)t^n \right)^2 \left(\sum_n L_n(1,y) t^n\right).
\end{eqnarray}
$G_2$ is singular at $t=e^{-\kappa(a)}$ and at $t=t_L(y)$ where $t_L(y)$ is the singular point of $\sum_n L_n(1,y) t^n$.
Clearly, when 
$y \ge 1$, $L_n(1,y)  \le C_n(1,y)$ so $t_L(y) \ge e^{-\lambda(y)}$.  Hence, for all three cases we see
that $\limsup_{n\to\infty}  {\scriptsize\frac{1}{n}} \log C_n(a,y)$ is bounded above by $\max[\kappa(a), \lambda(y)]$ which proves the Lemma.
\qed

\begin{theo}
When $a \ge 1$ and $y \ge 1$  
$$\psi(a,y)\equiv \lim_{n\to\infty}  {\scriptsize\frac{1}{n}} \log C_n(a,y) = \max [\kappa (a), \lambda(y)] =\psi^+(a,y).$$
\end{theo}
\Pr  This follows immediately from Lemma 1 and Lemma 2. \qed

\section{Walks repelled from the surface}
\label{sec:SAWrepel}

In this section we investigate the situation where the walk is repelled from  the surface
so that $a < 1$.  When $y=1$ we know that $\lim_{n\to\infty}  {\scriptsize\frac{1}{n}} \log C_n(a,1) = \kappa(a) = \log \mu_d$
for all $a \le 1$ \cite{HTW}.    In this section we extend this result to general values of $y$.  We first need a Lemma.

\begin{lemm}
 $$C_n(1,y) = \hbox{\large{$\frac{1}{y}$}} C_{n+1}(0,y)$$
 for all values of $y > 0$.
\end{lemm}
\Pr 
Every walk with $n$ steps and span $s$ can be converted to a walk with $n+1$ steps, span $s+1$ with only
one vertex in the hyperplane $x_d=0$.  This can be accomplished by translating the walk unit distance
in the positive $x_d$-direction and adding an edge to connect the walk to the origin.  This construction can be reversed
so that 
\begin{equation}
\sum_v c_n(v,s) = c_{n+1}(0,s+1).
\end{equation} 
Multiplying by $y^s$ and summing over $s$ gives
\begin{equation}
C_n(1,y) = \sum_{v,s} c_n(v,s) y^s = \sum_s c_{n+1} (0,s+1)y^s = y^{-1} C_{n+1} (0,y)
\end{equation}
and this proves the Lemma.
\qed

This is the key to proving the following theorem.
\begin{theo}
For $a \le 1$ and for all values of $y > 0$
$$\psi(a,y) = \lim_{n\to\infty}  {\scriptsize\frac{1}{n}} \log C_n(a,y) = \psi^+(a,y).$$
\end{theo}
\Pr  Fix $y$.  By monotonicity 
\begin{equation}
C_n(0,y) \le C_n(a,y) \le  C_n(1,y)
\end{equation}
for all $a \le 1$.  Take logarithms, divide by $n$ and let $n$ go to infinity.  Then using Lemma
3 we have 
\begin{equation}
\lim_{n\to\infty}  {\scriptsize\frac{1}{n}} \log C_n(a,y) = \lim_{n\to\infty}  {\scriptsize\frac{1}{n}} \log C_n(1,y)
\end{equation}
for all $a \le 1$ and for all values of $y$.  In \cite{Rensburg2016} it was shown that 
$\lim_{n\to\infty}  {\scriptsize\frac{1}{n}} \log C_n(1,y) = \psi^+(1,y)$ and in \cite{Rensburg2013} it was shown
that $\psi^+(a,y) = \psi^+(1,y)$ for all $a\le 1$ and for all $y$.  
Hence 
\begin{equation}
\fl \quad \quad
\psi(a,y) = \lim_{n\to\infty}  {\scriptsize\frac{1}{n}} \log C_n(a,y) = \lim_{n\to\infty}  {\scriptsize\frac{1}{n}} \log C_n(1,y) = \psi^+(1,y) = \psi^+(a,y)
\end{equation}
for all $a \le 1$ and for all $y$, which proves the Theorem.
\qed

In particular this means that $\psi(a,y) = \log \mu_d$ 
for $a \le 1$ and $y \le 1$.  For $a \le 1$ and $y \ge 1$ 
 $\psi(a,y) = \lambda (y)$.

\section{Walks pushed towards the surface}
\label{sec:SAWSpushed}

In this section we look at the situation where the force is directed towards the 
surface, $y < 1$, and the walk is attracted to the surface, $a > 1$.  We shall need several lemmas
about walks with fixed span.  One can think of the walk being confined to a slab, interacting
with one wall of the slab and having at least one vertex in the other wall.  Let
\begin{equation}
S_n^w(a) = \sum_v c_n(v,w) a^v
\end{equation}
so that $S_n^w(a)$ is the partition function for these walks with span $w$.

\begin{lemm}
The limit 
$$\lim_{n\to\infty}  {\scriptsize\frac{1}{n}} \log S_n^w(a) \equiv \kappa_w(a)$$
exists for all $a>0$.
\end{lemm}
\Pr
The proof is an easy adaptation of the results in Section 2 of \cite{HTW}.  It proceeds by concatenating unfolded walks 
with span $w$ and using most popular class arguments. \qed

\begin{lemm}
$\kappa_w(a)$ is a monotone non-decreasing function of $w$ so that $\kappa_{w+1}(a) \ge \kappa_w(a)$.
\end{lemm}
\Pr
Each walk counted by $S_n^w(a)$ has at least one vertex in $x_d=w$.  Suppose the vertex of degree 1 is in $x_d=w$.  Then add an edge in the $x_d$-direction and an edge in the $x_1$-direction to give a walk with $n+2$ edges and span $w+1$ counted by
$S_{n+2}^{w+1}(a)$.  If the vertex of degree 1 is not in $x_d=w$ then there must be an edge of the walk in 
$x_d=w$.  If there is more than one edge, take the one with lexicographically largest mid-point.  Translate this edge 
unit distance in the $x_d$-direction so that it lies in $x_d=w+1$ and add two edges to rejoin it to the remainder of the walk.  This 
gives a walk counted by $S_{n+2}^{w+1}(a)$. This means that $S_n^{w}(a) \le S_{n+2}^{w+1}(a)$.  Taking logarithms, dividing by 
$n$ and letting $n \to \infty$ completes the proof.
\qed

\begin{lemm}
The free energy of walks in a slab of width $W$ is identical to that of walks with span equal to $W$.
\end{lemm}
\Pr
Let $\widehat{S}_n^W(a) = \sum_{w \le W} S_n^w(a)$ be the partition function of self-avoiding walks with $n$ steps and 
\emph{maximum} span equal to $W$, so that these walks will fit in a slab of width $W$.  We know that
$S_n^w(a) = e^{n\kappa_w(a) +o(n)}$ so 
\begin{equation}
\fl \quad \quad
S_n^W(a) \le \widehat{S}_n^W(a) = \sum_{w \le W}  e^{n\kappa_w(a) +o(n)} \le (W+1) e^{n\kappa_W(a) +o(n)} = e^{n\kappa_W(a) + o(n)}.
\end{equation}
Hence $\lim_{n\to\infty}  {\scriptsize\frac{1}{n}} \log \widehat{S}_n^W(a) = \kappa_W(a)$.
\qed

\begin{figure}[t]
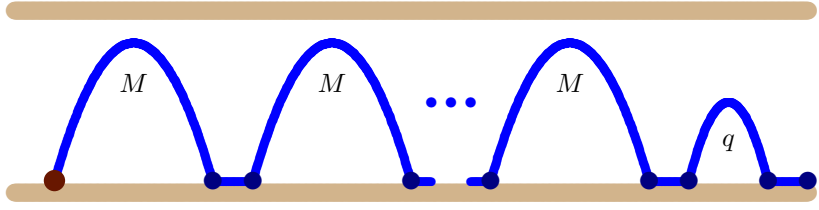

\beginpicture
\setcoordinatesystem units <1.5pt,1.5pt>
\setplotarea x from -100 to 150, y from -10 to 50
\setplotarea x from -50 to 150, y from -10 to 50

\color{Tan}
\setplotsymbol ({\LARGE$\bullet$})
\plot -50 -3 150 -3 /
\plot -50 43 150 43 /

\setquadratic
\color{Blue}
\setplotsymbol ({\footnotesize$\bullet$})
\plot -40 0 -20 35 0 0  /
\plot 10 0 30 35 50 0 /
\multiput {$\bullet$} at 55 20 60 20 65 20 /
\plot 70 0 90 35 110 0 /
\plot 120 0 130 20 140 0 /
\setlinear
\plot 0 0 10 0 / \plot 50 0 55 0 / \plot 65 0 70 0 / \plot 110 0 120 0 / \plot 140 0 150 0 /
\color{NavyBlue}
\multiput {\LARGE$\bullet$} at 0 0 10 0 50 0 70 0 110 0 120 0 140 0 150 0 /

\setlinear
\color{Sepia}
\put {\huge$\bullet$} at -40 0 
\color{black} \normalcolor
\multiput {$M$} at -20 25 90 25 30 25 /
\put {$q$} at 130 10

\endpicture

\caption{Concatenating $p$ unfolded loops with $M$ steps and an unfolded loop with $q$ steps.}
\label{fig:loopconcat}
\end{figure}

\begin{lemm}
$$\lim_{w\to\infty} \kappa_w(a) = \sup_w \kappa_w(a) = \kappa(a).$$
\end{lemm}
\Pr
Clearly $\lim_{n\to \infty}  {\scriptsize\frac{1}{n}} \log \widehat{S}_n^w(a) \le \kappa (a)$ and we know that
$\kappa_w(a)$ is a monotone non-decreasing function of $w$.
If we concatenate a set of unfolded loops each with $M$ steps we obtain a subset of walks that
fit in a slab of width $M$.  See Figure \ref{fig:loopconcat} for a sketch.  Fix $M$ and write $n=Mp+q$, $0 \le q < M$.  Concatenate 
$p$ unfolded loops of $M$ steps and a final unfolded loop of $q$ steps.  This gives the inequality
\begin{equation}
L_M^{\ddagger}(a,1)^p L_q^{\ddagger}(a,1) \le \widehat{S}_n^M(a).
\end{equation}
Take logarithms and divide by $n$ giving
\begin{equation}\left(\frac{n-q}{Mn}\right) \log L_M^{\ddagger}(a,1) + \frac{1}{n} \log L_q^{\ddagger}(a,1) \le 
\frac{1}{n} \log \widehat{S}_n^M(a).
\end{equation}
Let $n \to \infty$ giving
\begin{equation}
\frac{1}{M} \log L_M^{\ddagger} (a,1) \le \kappa_M(a)
\end{equation}
and then letting $M \to \infty$ we have $\kappa (a) \le \lim_{M\to\infty} \kappa_M(a)$
which proves the Theorem.
\qed

\begin{figure}[h]
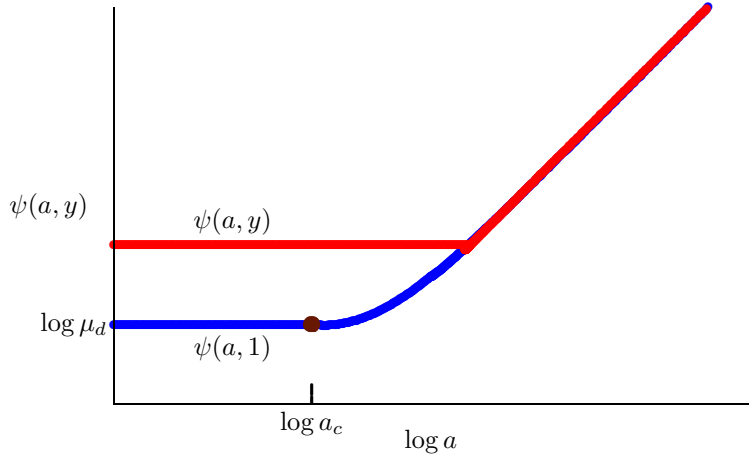

\beginpicture
\setcoordinatesystem units <1.5pt,1.5pt>
\setplotarea x from -160 to 90, y from -10 to 110
\setplotarea x from -100 to 60, y from 0 to 100

\setplotsymbol ({\footnotesize$\bullet$})
\color{Blue}
\plot -100 20 -50 20 /
\setquadratic
\plot -50 20 -36 22 -20 32 0 50 50 100  /
\color{Red}
\setlinear
\plot -100 40 -10 40 /
\setquadratic
\plot -10 40 0 50 50 100  /
\setlinear

\color{black}
\setplotsymbol ({.})
\axis bottom  label $\log a$ /
\axis left label $\psi(a,y)$ /

\plot -50 0 -50 5 /
\put {$\log a_c$} at -50 -5
\put {$\log \mu_d$} at -110 20 

\put {$\psi(a,1)$} at -70 14
\put {$\psi(a,y)$} at -70 46

\color{Sepia}
\put {$\huge\bullet$} at -50 20 
\setplotsymbol ({\footnotesize$\bullet$})
\circulararc 360 degrees from -49 20 center at -50 20
\color{black} \normalcolor
\endpicture

\caption{The limiting free energy $\psi(a,y)$ plotted against $\log a$ for $y=1$ (bottom curve)
and for $y>1$ (top curve).  There is an adsorption transition when $y=1$ at $a=a_c$.
In the case that $y>1$ the adsorption transition is present at a critical point larger
than $a_c$, and the transition is first order.  Notice that for $y<1$, 
$\psi(a,y) = \psi(a,1)$ as shown in theorem 3.}
\label{fig:walkFE}
\end{figure}

We now turn to the main result of this section which we state as the following theorem.

\begin{theo}
For all $a \ge 1$ and $y \le 1$ 
$$\psi(a,y) = \lim_{n\to\infty}  {\scriptsize\frac{1}{n}} \log C_n(a,y) = \kappa(a).$$
\end{theo}
\Pr 
The same walks are counted by $C_n(a,y)$ and by $C_n^+(a,y)$ but with different weights (in general).
The span is at least as large as the height of the last vertex so, for $y \le 1$,
\begin{equation}
C_n(a,y) \le C_n^+(a,y).
\end{equation}
Hence 
\begin{equation} \limsup_{n\to\infty}  {\scriptsize\frac{1}{n}} \log C_n(a,y) \le \psi^+(a,y) = \kappa (a)
\label{eqn:ub}
\end{equation}
for $y \le 1$.
To obtain a bound in the opposite direction we note that, for any $y\le 1$,
\begin{equation}
C_n(a,y) \ge y^w \sum_v c_n(v,w) a^v = y^w S_n^w(a).
\end{equation}
Therefore
 \begin{equation}
\liminf_{n\to\infty}  {\scriptsize\frac{1}{n}} \log C_n(a,y) \ge \kappa_w(a)
\label{eqn:lb}
\end{equation}
for all $w$.  But $\sup_w \kappa_w(a) = \kappa(a)$ and so, using (\ref{eqn:ub}) and (\ref{eqn:lb}),
\begin{equation} 
\lim_{n\to\infty}  {\scriptsize\frac{1}{n}} \log C_n(a,y) = \kappa (a)
\end{equation}
for all $y \le 1$.
\qed

\section{Discussion}
\label{sec:discussion}

We have considered the problem of a self-avoiding walk on the hypercubic lattice 
${\mathbb Z}^d$, starting at the origin and confined between two parallel hyperplanes, one through
the origin and one being the hyperplane containing the top set of vertices.  The walk interacts with
the plane through the origin and can adsorb at that plane.  A force is applied at the other confining plane,
perpendicular to the adsorbing plane and either directed away from the adsorbing  plane or towards it.
In the first case the (adsorbed) walk is pulled off the adsorbing plane and in the second case it is 
pushed towards this plane.   In both cases we prove that the free energy is identical
to the case where the force is applied at the last vertex of the walk.  This shows that our 
criterion for determining the critical force - temperature curve is identical for the 
two modes of pulling, and the response to pushing the walk is the same in the two
cases in the thermodynamic limit. In Figure \ref{fig:walkFE} we sketch the expected form of the free
energy as a function of $\log a$ for $y=1$  (no force) where the phase transition is expected
to be second order, and for $y > 1$ where the transition is known to be first order \cite{Guttmann2014}.

\section*{Acknowledgement}
This research was partially supported by NSERC of Canada.

\end{document}